\documentclass[%
floatfix,
 reprint,
superscriptaddress,
 amsmath,amssymb,
 aps,
 onecolumn
]{revtex4-2}

\usepackage{graphicx}
\usepackage{hyperref}
\usepackage{float}
\usepackage{subcaption}
\usepackage[utf8]{inputenc}
\usepackage[T1]{fontenc}
\usepackage{amsmath}
\usepackage{cleveref}
\usepackage{xcolor}
\usepackage{lineno}

\begin{document}


\title{Insights and caveats from mining local and global temporal motifs in cryptocurrency transaction networks}

\author{Naomi Arnold}
\email[]{naomialicearnold@gmail.com}
\affiliation{Northeastern University London, Network Science Institute, London, E1W 1LP, United Kingdom}
\affiliation{%
Queen Mary University of London, School of Electronic Engineering and Computer Science, London, E1 4NS, United Kingdom
}%
\author{Peijie Zhong}
\affiliation{%
Queen Mary University of London, School of Electronic Engineering and Computer Science, London, E1 4NS, United Kingdom
}%
\author{Cheick Tidiane Ba}
\affiliation{%
Queen Mary University of London, School of Electronic Engineering and Computer Science, London, E1 4NS, United Kingdom
}%
\affiliation{University of Milan, Department of Computer Science, Via Giovanni Celoria, 18, 20133 Milano MI, Italy}
\author{Ben Steer}
\affiliation{Pometry Ltd, 69 Wilson St, London, EC2A 2BB, United Kingdom}
\author{Raul Mondragon}
\affiliation{%
Queen Mary University of London, School of Electronic Engineering and Computer Science, London, E1 4NS, United Kingdom
}%
\author{Felix Cuadrado}
\affiliation{Universidad Politecnica de Madrid, School of Telecommunications Engineering, Av. Complutense, 30, Moncloa - Aravaca, 28040 Madrid, Spain}
\author{Renaud Lambiotte}
\affiliation{University of Oxford, Mathematical Institute, Oxford, OX2 6GG, United Kingdom}
\affiliation{Pometry Ltd, 69 Wilson St, London, EC2A 2BB, United Kingdom}
\author{Richard G. Clegg}
\affiliation{%
Queen Mary University of London, School of Electronic Engineering and Computer Science, London, E1 4NS, United Kingdom
}%

\begin{abstract}
Distributed ledger technologies have opened up a wealth of fine-grained transaction data from cryptocurrencies like Bitcoin and Ethereum. This allows research into problems like anomaly detection, anti-money laundering, pattern mining and activity clustering (where data from traditional currencies is rarely available). The formalism of temporal networks offers a natural way of representing this data and offers access to a wealth of metrics and models. However, the large scale of the data presents a challenge using standard graph analysis techniques. We use temporal motifs to analyse two Bitcoin datasets and one NFT dataset, using sequences of three transactions and up to three users. We show that the commonly used technique of simply counting temporal motifs over all users and all time can give misleading conclusions. Here we also study the motifs contributed by each user and discover that the motif distribution is heavy-tailed and that the key players have diverse motif signatures. We study the motifs that occur in different time periods and find events and anomalous activity that cannot be seen just by a count on the whole dataset. Studying motif completion time reveals dynamics driven by human behaviour as well as algorithmic behaviour.
\end{abstract}

\maketitle
\section{Introduction}
Digital assets have a growing importance in the modern world with cryptocurrencies and Non-Fungible Tokens (NFTs) being important examples: cryptocurrencies are analogous to digital wealth and NFTs are analogous to digital ownership. An advantage for researchers is that the transfer is recorded, almost always publicly, on a blockchain giving valuable data-sets and opportunities to bring sophisticated analysis tools to bear. This enables us to gain insight into the nature of these ecosystems. An additional challenge in this kind of analysis is that these systems are relatively young; Bitcoin (BTC) was created in 2008~\cite{nakamoto2008bitcoin}, NFTs in 2014 and hence they change rapidly over time. The systems can be viewed as a set of exchanges between a pair of users with each exchange having a timestamp and an amount. It is natural to view and analyse such systems as weighted temporal graphs, where nodes are users and a directed edge represents a transaction from the source to the destination node.

The paper shows that temporal motifs are a powerful tool for analysis of transaction networks. We extend the work of~\cite{paranjape2017motifs} by providing methods that allow temporal motifs to be examined locally (that is, counting the number and type of temporal motifs centred around each node) and temporally (that is showing how the number and type of motifs evolve over time). These extensions rely on integrating our algorithms with our Raphtory software~\cite{steer2024raphtory} that allows efficient parallel computation of temporal structures in networks. This analysis would not be possible using the software developed in~\cite{paranjape2017motifs}. By taking a deep dive into datasets centred on cryptocurrency we show how these tools can be used in practice.
The paper also presents a number of caveats: (i) Traditional assumptions about temporal ordering within a motif do not hold precisely with transactions within a single block (we show that the effects of this on our results are minor but this might not be the case for all possible analysis). (ii) Motif counts of the transaction graph as a whole might give a distribution of motifs that places undue attention on events limited to a single node or a small number of nodes and (iii) limited to a short span of time. Using motif analysis on these networks requires researchers to investigate the distribution both across time and across individual nodes. We show how the tools and techniques developed for this research can achieve this. 

\section{Background}
\subsection{Temporal motif definition}\label{sec:motif-definition}

Motifs~\cite{alon2002network,yaveroglu2014revealing} are small commonly occurring subgraphs in a network, which can be thought of as measuring the organisation of that network at a small scale. The study of motifs has been of great importance in the fields of biology~\cite{vazquez2004topological}, anatomy~\cite{sporns2004motifs} and social science~\cite{holland1976local}. Temporal motifs~\cite{paranjape2017motifs,kovanen2011temporal} add a requirement that the motif takes place in a set order and within a certain timeframe.

\begin{figure*}[htbp]
    \centering
    \subcaptionbox{}{\includegraphics[width=0.20\linewidth]{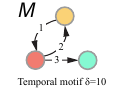}} \hfill
    \subcaptionbox{}{\includegraphics[width=0.20\linewidth]{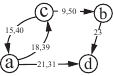}} \hfill
    \subcaptionbox{}{\includegraphics[width=0.20\linewidth]{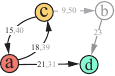}} \hfill
    \subcaptionbox{}{\includegraphics[width=0.20\linewidth]{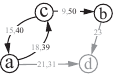}}
    \caption{An example of extracting a particular temporal motif from a temporal graph. \textbf{(a)} is an example of $\delta$-temporal motif $M$ with a given $\delta=10$; \textbf{(b)} is a temporal graph with edges appearing at the times shown on each edge; \textbf{(c)} shows an instance of $\delta$-temporal motifs in the temporal graph; \textbf{(d)} is not a $\delta$-temporal motif because the difference between the timestamp of the first temporal edge and the timestamp of the last temporal edge exceeds the given $\delta$.}
    \label{fig:temporal_motif}
\end{figure*}
We begin by defining what we mean by a temporal motif following Paranjape \textit{et al.}~\cite{paranjape2017motifs}. Let the tuple $(u,v,t)$ define a transaction from user $u$ to user $v$ at time $t$, and let $\delta > 0$ represent a time interval. Define a $k$-node $l$-edge temporal motif as a sequence of edges $M = (u_1, v_1, t_1), \dots , (u_l, v_l, t_l)$ such that $t_1 < t_2 < \dots < t_l$ and $t_l - t_1 \leq \delta$, and the graph induced by these edges is connected with $k$ nodes. $M' = (u'_1, v'_1, t'_1), \dots , (u'_l, v'_l, t'_l)$ is an instance of motif $M$ if there is a mapping $u'_i \mapsto u_i, v'_i \mapsto v_i$ for $1 \leq i \leq l$ and the same constraints as $M$ are obeyed. \Cref{fig:temporal_motif} shows a temporal motif $M$, a timestamped graph, an example instance and non-instance of $M$ for a given $\delta$. In this paper we consider three-edge up-to-three-node temporal motifs.

\begin{figure*}[htbp]
    \centering
    \includegraphics[width=0.5\linewidth]{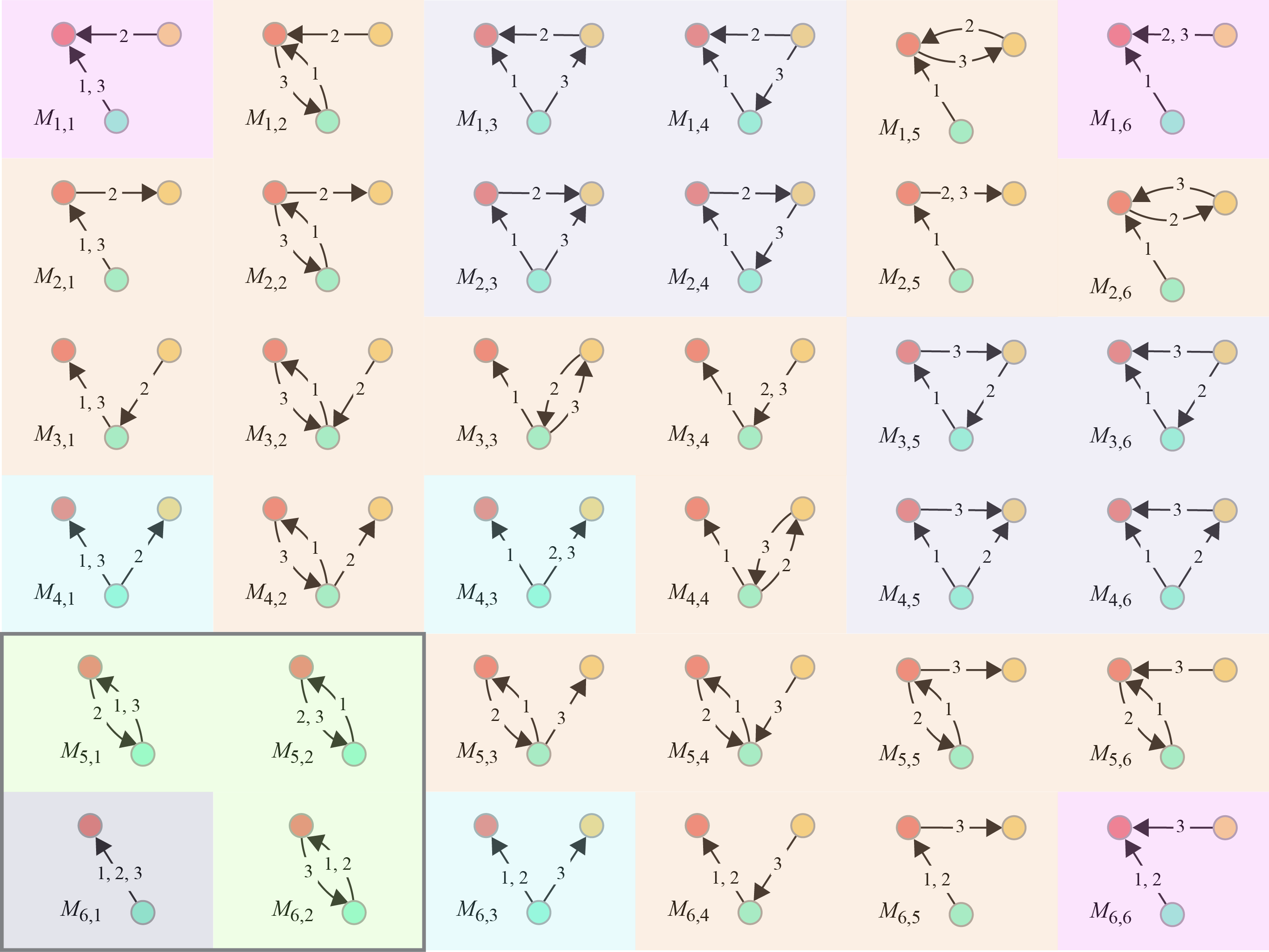}
    \caption{Possible three-edge up-to-three node motifs considered in this work, using the same enumeration as in Paranjape et al~\cite{paranjape2017motifs} with added colouring to indicate sub-types. For local motif counts, we consider an additional four two-node motifs which are exactly those pictured but with directions reversed (the direction becomes important when counting from the perspective of a node).}
    \label{fig:motif-types}
\end{figure*}

When considering the global motif count there are thirty six possible motifs of this type (pictured in~\cref{fig:motif-types}). We divide them here into six classes according to the number of nodes, the subgraph formed by aggregating the motif's edges, and the direction of the transactions involved, along with their colour in~\cref{fig:motif-types}:
(i) Two-node motifs, all transactions in the same direction (dark grey); (ii) two-node motifs with transactions in mixed directions (green); (iii) a three node ``star" (a centre node with two nodes connected to it) with all transactions incoming to the centre (pink); (iv) a three node ``star" with all transactions outgoing from the centre (blue); (v) a three node star with mixed transaction directions (orange) and finally (vi) triangular motifs (light grey) which can have transactions in a cyclic direction (e.g. $M_{2,4}$ and $M_{3,5}$) or mixed directions. For a global motif count these are the only possible motifs with three edges and up to three nodes. \Cref{fig:motif-types} is shaded to distinguish these motifs. With a local motif count then we can distinguish some extra motif types. For example, three transactions in the same direction between the same pair of nodes (the lower left box in~\cref{fig:motif-types}) would be seen as ``all outgoing" from the point of view of one node but ``all incoming" from the point of view of the other. The two node motifs that can be distinguished from a local node (but not a global perspective) are surrounded by a box in~\cref{fig:motif-types}. 

\subsection{Related work}

Temporal motifs have been used for studying financial transactions~\cite{liu2023temporal}, call networks~\cite{kovanen2013temporal}, email interactions~\cite{barnes2023temporal} and patent networks~\cite{liu2022temporal}. In the context of a \emph{transaction networks}, temporal motifs can give us an insight into the pattern of trade. For example, if we often see the motif: A sends to B then B sends to C, it implies node B is often a ``middleman" in the trade but if we often see the motif: A sends to B then A sends to C it suggests that node A may be a buyer (sending virtual currency for goods or services) or an on-ramp (trading virtual currency for real world currency). Motif counting has been used within cryptocurrency networks to detect addresses associated with mixing services~\cite{wu2021detecting} (services designed to obfuscate the intended sender and recipient of a transaction) or for detecting phishing scams in Ethereum~\cite{wang2023detecting} where authors posed this as a classification problem. These authors found that the inclusion of temporal motif features improved their classification results beyond using basic transaction features. Motifs have also been used to understand the longer-term formation of the socio-economic network of traders, using \emph{graph evolution rules} to highlight how different structures like closed triangles emerge between traders~\cite{galdeman2022disentangling,ba2023characterizing}. More generally, static motifs have been used as a descriptive graph metric for comparing cryptocurrency networks to other social networks~\cite{lee2020measurements}, as features for a token price prediction pipeline~\cite{chen2019deep} and to investigate the rise of smart contracts in Ethereum, typically involving groups of pre-programmed transactions~\cite{wu2023know}. Motifs are also related to \emph{cycles} which have been used to study wash trading and market manipulation in NFT markets by our team~\cite{yousaf2023non} and others~\cite{von2022nft}.

Most of the literature mentioned uses temporal motifs for a specific task such as detecting a type of entity or anomalous activity. In this paper, we step back and take a more explorative approach, highlighting the insights that can be gained using temporal motifs into cryptocurrency systems as a whole as well as individuals within them, making use of three different datasets. Two of the datasets are BTC transactions centred around different ``dark web marketplaces"~\cite{nadini2022emergence,elbahrawy2020collective} (services on the dark web that exist to facilitate the exchange of goods and services, often illegal) and a third is a collection of NFT sales. We use temporal motifs with three edges and a maximum of three nodes~\cite{paranjape2017motifs} but with two further steps: (i) we disaggregate the motifs by user, that is to obtain a \emph{local} motif count for every vertex in the graph and (ii) we study the whole graph motif counts over different times and different timescales. We find that counting local motifs produces a very different view of the network from simply looking at node degree or weighted node degree. It gives prominence in the analysis to a different set of nodes (the highest degree node does not have the largest motif count) and the motif count over time shows a very different trend to the transaction count over time. It also allows insight into the nature of and importance of nodes in the network that might be overlooked by more traditional temporal graph methods. 

\subsection{Datasets}
Three cryptocurrency datasets are used in this paper. The Alphabay and Hydra datasets are bitcoin transactions centred around the Alphabay and Hydra dark web marketplaces respectively. These marketplaces act as an escrow service, such that a typical (successful) purchase on these platforms involves a transaction from the buyer to the market node followed by a transaction from the market node to the seller after the goods have been confirmed as received. They include transactions up to two hops away from the marketplace nodes (i.e. the transactions of all users who have bought or sold on these markets); this includes transactions to and from the marketplace itself. The NFT dataset is formed of transactions representing the purchase of an NFT, usually but not always in Ethereum. Unlike the dark web marketplace purchases, NFT purchases on a marketplace like OpenSea are executed using a smart contract initiating a direct transaction from buyer to seller and a small platform fee is taken by OpenSea. We do not observe transactions representing a platform fee, just the buyer to seller transactions.

In Bitcoin, one person tends to have multiple different wallets (for example, a basic transaction involving user $u$ sending an amount to user $v$ will involve an additional change wallet belonging to user $u$). The Alphabay and Hydra datasets are therefore pre-processed by Chainalysis Inc. such that wallets who are believed to be associated with the same person are merged into one entity in the graph. The NFT purchases network is not preprocessed in this way. In the NFT network the direction of the transfer is the direction in which money is sent, not the direction in which the NFT is sent, this makes it consistent with the other two datasets. Part of the nature of blockchain systems is that transactions are grouped in ``blocks". If two transactions are within the same block it is not certain which order they came in, although it might be argued that a transaction at the end of the block is highly likely to occur after one at the beginning. This issue is discussed in depth in~\cref{sec:timeorder}.

In all datasets, nodes represent wallets (potentially a group of wallets as discussed above for Alphabay and Hydra); a directed edge from node $u$ to node $v$ at time $t$ represents a transaction from the user represented by node $u$ to the user represented by node $v$. \Cref{demo-table} gives some statistics on the number of nodes, the number of edges (the number of ordered pairs $(u,v)$ such that there is at least one transaction from $u$ to $v$) and the total number of transactions.

\begin{table*}[htbp]
\begin{center}
\begin{tabular}{c  c  c  c  c } 
\hline
\textbf{Dataset} & \textbf{Time range(mm/yy)} & \textbf{Number of nodes} &\textbf{Number of directed edges} & \textbf{Number of transactions} \\ [0.5ex] 
 \hline
 Alphabay &  09/2010--12/2020 & 11,021,692 & 15,890,235 & 33,588,967 \\ 
 Hydra & 07/2010--12/2020 & 20,137,164 & 35,765,997 & 69,811,848\\
 NFT & 11/2017--04/2021 & 532,945 & 2,991,602 & 6,071,027 \\
 \hline
\end{tabular}
\caption{\label{demo-table}Dataset summary. Time range (earliest to latest recorded transaction), number of nodes, number of directed edges and number of transactions. Note the distinction between \emph{edges}, the number of unique pairs $(u,v)$ such that there is at least one transaction from $u$ to $v$, and transactions. }
\end{center}
\end{table*}

Throughout the paper we will refer to the data sets as Alphabay, Hydra and NFT. It's important to remember that Alphabay and Hydra can be viewed as different subsets of the same underlying dataset (the global bitcoin transfer network). 

\section{Results}
\subsection{Motifs carry different information to (weighted) node degree}
\begin{figure*}[htbp]
    \centering
    \begin{subfigure}[b]{0.32\textwidth}
        \centering \includegraphics[width=\linewidth]{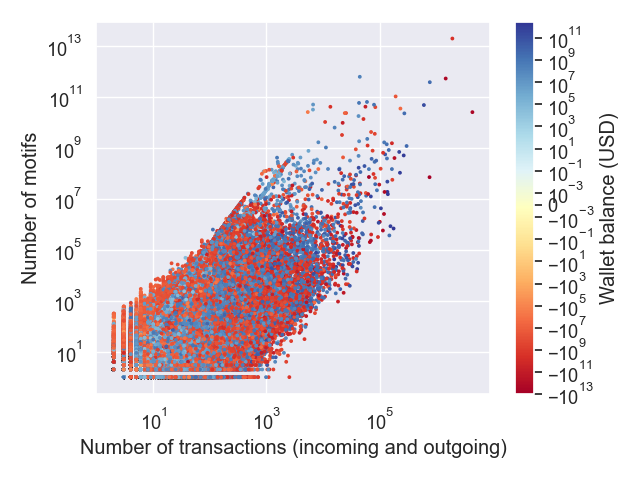}
        \caption{Alphabay}
        \label{fig:alpha-vs-transactions}
    \end{subfigure}
    \hfill
    \begin{subfigure}[b]{0.32\textwidth}
        \centering \includegraphics[width=\linewidth]{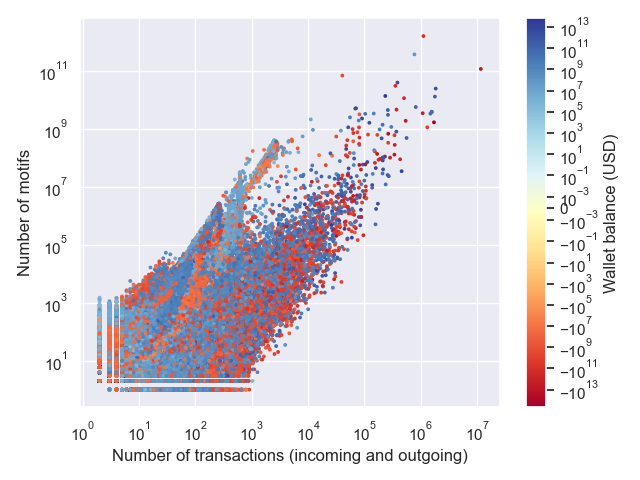}
        \caption{Hydra}\label{fig:hydra-vs-transactions}
    \end{subfigure}
    \hfill
    \begin{subfigure}[b]{0.32\textwidth}
        \centering \includegraphics[width=\linewidth]{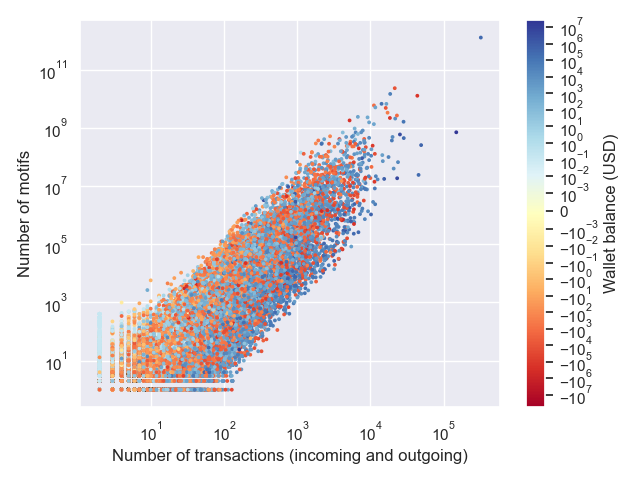}
        \caption{NFTs}\label{fig:nft-vs-transactions}
    \end{subfigure}
    \caption{Local motif counts for nodes in the Alphabay network with $\delta$ set to one hour. The plot shows total motif count versus total transactions (both incoming and outgoing). The nodes are coloured by their wallet balance (total incoming transaction value minus total outgoing transaction value). Note the scale is logarithmic in both axes and in the colouring.}
    \label{fig:alpha-motif-weight}
\end{figure*}
It is important to establish that motif counts are giving different information than simply looking at, for example, node degree. In~\cref{fig:alpha-motif-weight} for each node we plot the number of motifs (with $\delta$ of one hour for Alphabay/Hydra and one day for NFTs) against the number of transactions. We colour the node by the wallet balance, that is the amount in dollars coming in to that node minus the amount going out within the time period. An obvious feature is that while there is a clear correlation between the number of transactions and the number of motifs, this varies by many orders of magnitude; for example a node that is involved in 1,000 transactions may be part of fewer than 10 motifs but may be involved in as many as 100 billion. The node in~\cref{fig:alpha-vs-transactions} with the most transactions (corresponding to Alphabay itself) is not in the top ten for motif count. Besides the large fluctuations in the relations between number of transactions and number of motifs, one also observes non-trivial patterns, such as in~\cref{fig:hydra-vs-transactions}, which may be associated to the existence of different types of users in the system.
\subsection{Global motif counts show fanout patterns of trade}
\begin{figure*}[htbp]
    \centering
    \begin{subfigure}[b]{0.32\textwidth}
        \centering \includegraphics[width=\linewidth]{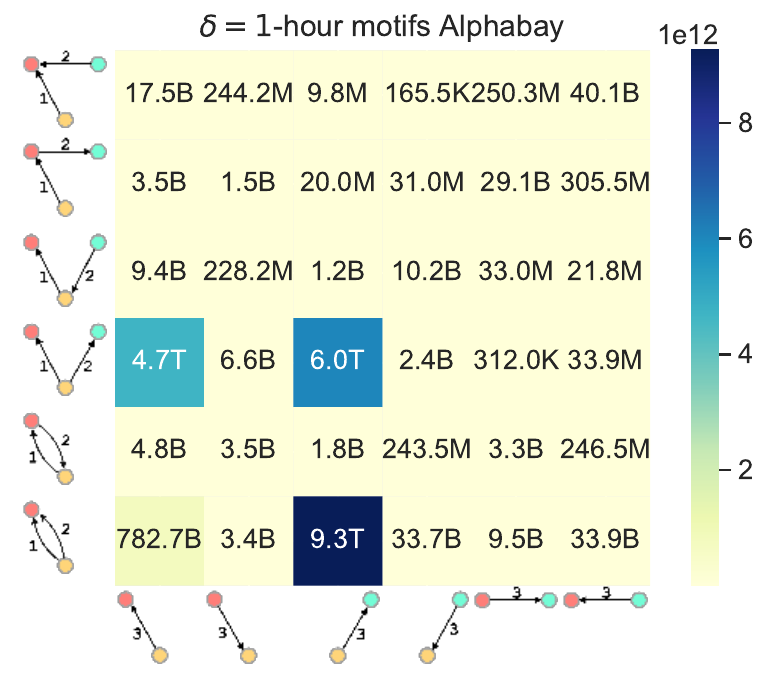}
        \caption{Alphabay, $\delta=1$ hour}
        \label{fig:alpha-1h-grid}
    \end{subfigure}
    \hfill
        \begin{subfigure}[b]{0.32\textwidth}
        \centering \includegraphics[width=\linewidth]{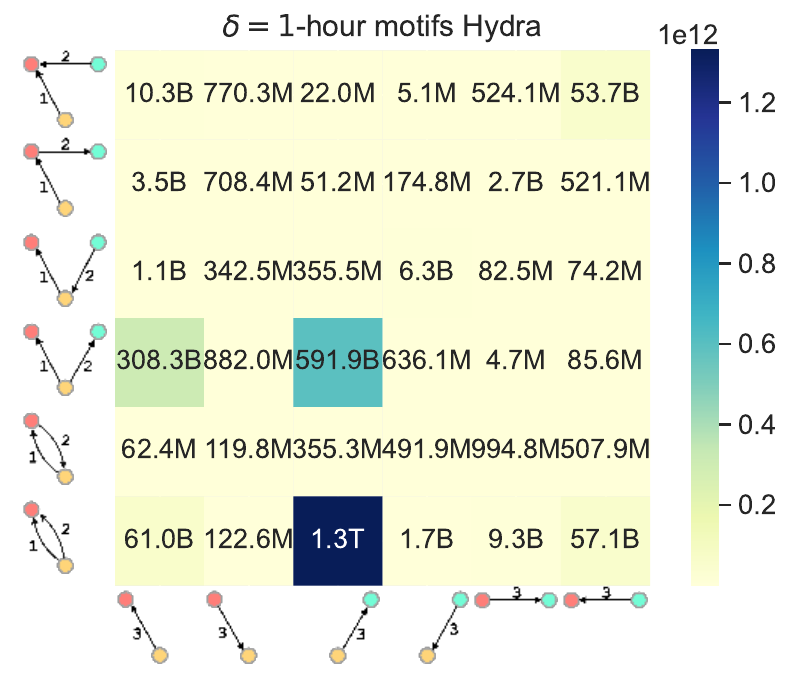}
        \caption{Hydra, $\delta=1$ hour}
        \label{fig:hydra-1h-grid}
    \end{subfigure}
    \hfill
    \begin{subfigure}[b]{0.32\textwidth}
        \centering \includegraphics[width=\linewidth]{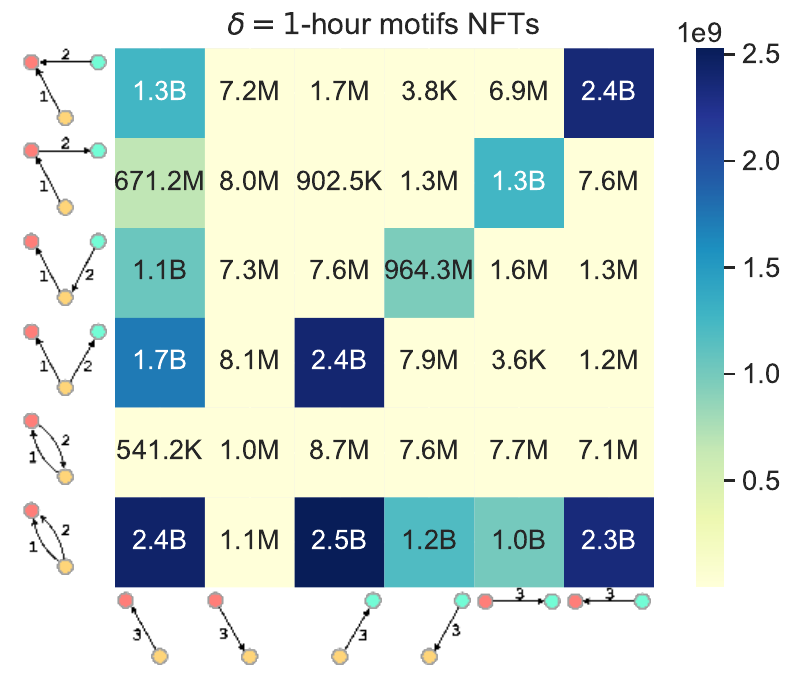}
        \caption{NFTs, $\delta=1$ day}
        \label{fig:nft-1d-grid}
    \end{subfigure}
    \hfill
    \begin{subfigure}[b]{0.32\textwidth}
        \centering \includegraphics[width=\linewidth]{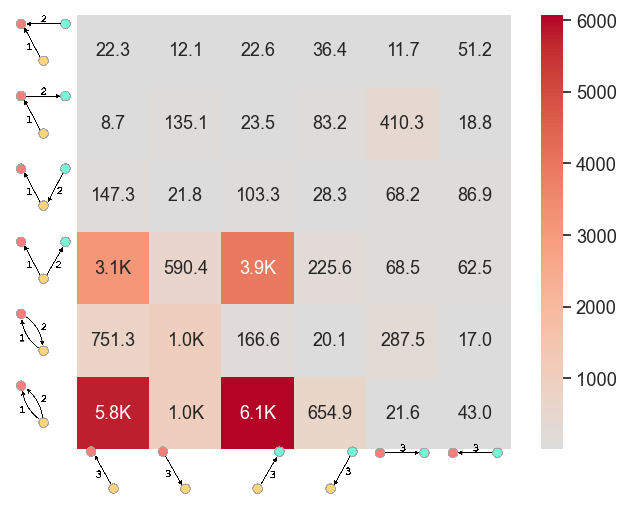}
        \caption{Alphabay vs time-shuffled, $\delta=1$ hour}
        \label{fig:alpha-1h-relative}
    \end{subfigure}
    \hfill
        \begin{subfigure}[b]{0.32\textwidth}
        \centering \includegraphics[width=\linewidth]{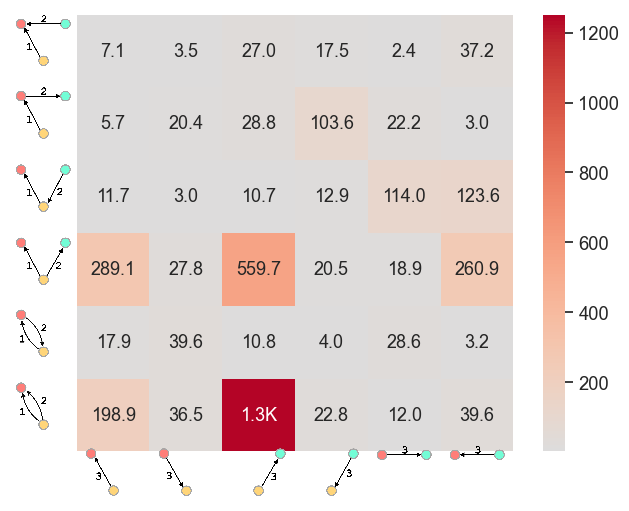}
        \caption{Hydra vs time-shuffled, $\delta=1$ hour}
        \label{fig:hydra-1h-relative}    
    \end{subfigure}
    \hfill
    \begin{subfigure}[b]{0.32\textwidth}
        \centering \includegraphics[width=\linewidth]{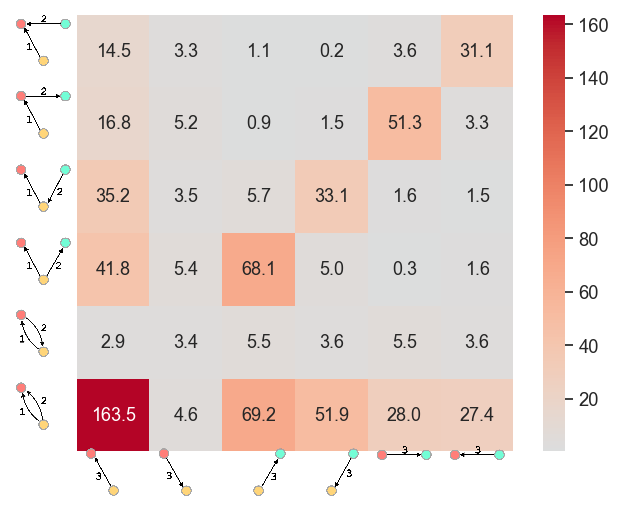}
        \caption{NFTs vs time-shuffled, $\delta=1$ day}
        \label{fig:nft-1d-relative}    
    \end{subfigure}
    \caption{Global motif counts for each dataset (top) compared with a null time shuffled model (bottom) where the lower graphs show the ratio of motif count in the unshuffled versus shuffled data. (See text for axes interpretation.)}
    \label{fig:motif-relative-grid}
\end{figure*}

In~\cref{fig:alpha-1h-grid,fig:hydra-1h-grid,fig:nft-1d-grid} we look at the global number of motifs for each dataset. A first thing to note is the huge variation in motif counts. The highest motif count for Alphabay is 9.3 trillion for an outgoing star motif whereas the smallest motif count is 164 thousand for a mixed triangular motif count. It is clear that some types of motif have extremely high prevalence in this data compared to others.  Still, one may wonder if these high counts of motifs occur by chance due (say) to an entity having a high number of outgoing transactions and the motifs are simply a product of a large number of edges between certain node pairs. To investigate this, we use a randomised reference model (see~\cref{sec:nullmodel}) which preserves the source and destination of all transactions but permutes the time at random, so that the overall activity rate and aggregate graph structure is preserved but the order of events is lost. \Cref{fig:alpha-1h-relative,fig:hydra-1h-relative,fig:nft-1d-relative} show the ratio between the motif counts in the original and shuffled network, showing that for Alphabay the four all-outgoing motifs are thousands of times overrepresented. All motifs are represented more than would be expected in the null reference model. Similar results hold for the Hydra dataset. An outbound pattern corresponds to an entity which is largely sending currency, perhaps buying physical goods, performing automated services or exchanging to some other (perhaps fiat) currency. 
For the NFT data we found $\delta$ of one day was a more appropriate time scale (again see~\cref{sec:timescale}). This may be because the NFT data set has a slower rate of arrival of edges. The NFT data set shows more diversity of motif types. The extreme difference in count magnitudes is still present but is reduced (the largest motif has a count of 1.6 billion for an outgoing star, the smallest, 471 thousand for a mixed direction triangle). While outbound star patterns are still the most common motifs, mixed star patterns with one node always inbound and one node always outbound are also common, perhaps a user selling one NFT to fund the purchase of another or buying an NFT and selling it shortly after. It is also worth noting that triangular motifs are rare in the NFT marketplace (again this is not unexpected, there is no particular reason three traders would wish to exchange NFTs between themselves) and in some cases the number of triangle motifs is lower than the chance results from the time-shuffled null model ($M_{1,4}$ and $M_{4,5}$). 

\textbf{A note on interpreting motif count plots:} In~\cref{fig:motif-relative-grid} the cell $i,j$ corresponds to the count  $M_{i,j}$ in~\cref{fig:motif-types}. However, the axes can be used to interpret which motif is being used without referring to that figure. The x and y axes show the directions and participants for the three transactions. The row (y-axis) corresponds to the first two transactions and the column (x-axis) corresponds to the third.  For example, the top right box is the motif count for transactions from yellow to red, then blue to red then blue to red a second time (a type of inbound star motif).

\subsection{Motif counts are dominated by a very small number of nodes}
\label{sec:dominated}
\begin{figure*}[htbp]
    \begin{subfigure}[t]{.32\textwidth}
        \centering
        \includegraphics[width=\linewidth]{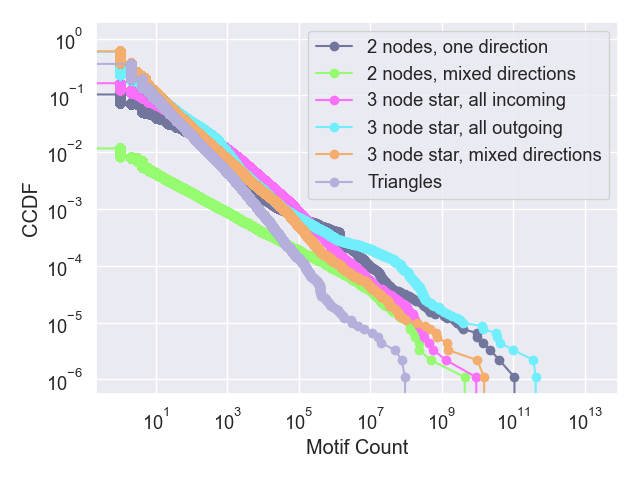}
        \subcaption{Alphabay ($\delta$=1 hour)}
        \label{fig:ccdf-alpha}
    \end{subfigure}
    \hfill
    \begin{subfigure}[t]{.32\textwidth}
        \centering
        \includegraphics[width=\linewidth]{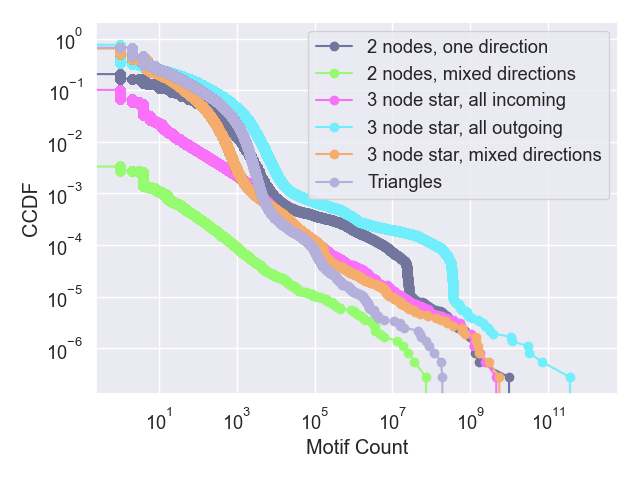}
        \subcaption{Hydra ($\delta$=1 hour)}
        \label{fig:ccdf-hydra}
    \end{subfigure}
    \hfill
    \begin{subfigure}[t]{.32\textwidth}
         \centering
         \includegraphics[width=\linewidth]{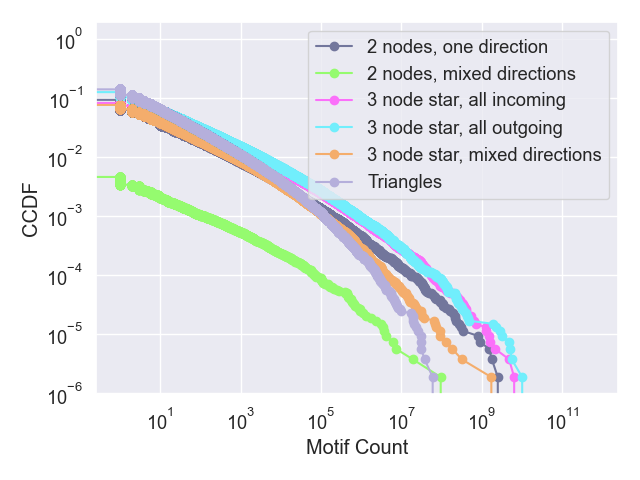}
         \subcaption{NFT ($\delta$ = 1 day)}
         \label{fig:ccdf-nft-1-day}
     \end{subfigure}
    \caption{Complementary cumulative distribution function of the motifs each node participates in, grouped into categories of similar motifs for Alphabay and Hydra and NFT. Both axes are log-scaled.}
    \label{fig:ccdf-all}
\end{figure*}

Given the magnitude of the motif counts found in the graph-wide counts, we next ask how these motifs are dispersed across the nodes in the network. \Cref{fig:ccdf-all} shows the distribution of motifs across the nodes in the network. For Alphabay we largely see straight lines in the log-log CCDF which is characteristic of power law behaviour and shows huge dominance by the nodes with the most motifs. In Hydra only the three-node star with all incoming edges and two-node same direction motifs are power-law, the others display sudden downward ``bumps'' (e.g. around $2x10^7$ motifs for 3 node star all outgoing). These ``bumps" indicate a group of nodes with extremely similar motif counts, one possible explanation for this is the presence of groups of nodes implementing similar automated behaviour.
For the NFT data~\Cref{fig:ccdf-nft-1-day}  the situation is similar but less extreme. The curves are smoother than in the Alphabay and Hydra and fall off towards higher motif counts. However, they still indicate a situation where the bulk of the count of motifs is with a small number of nodes. The two-node mixed direction motif has a much smaller count than any other in this data set. This is unsurprising as it would indicate two people buying NFTs from each other in quick succession. 

\subsection{Motifs are varied among the top 10 nodes}\label{sec:localmotifs}
\begin{figure*}[htbp]
    \centering
    \includegraphics[width=\linewidth]{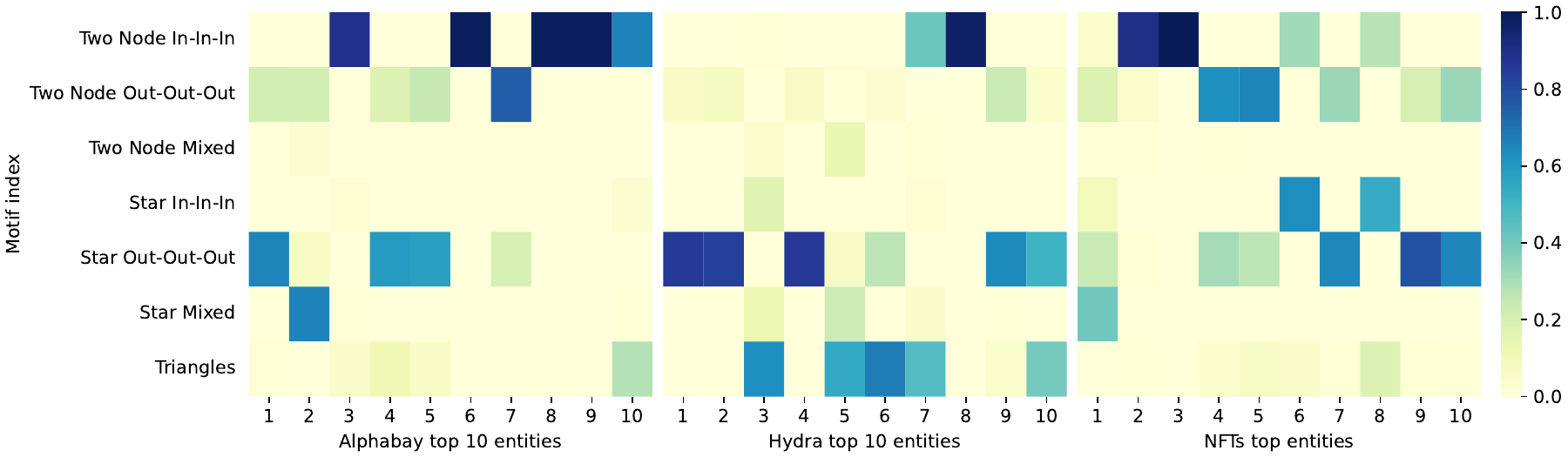}
\caption{Motif `signatures' among the top 10 entities in Alphabay, Hydra and NFTs. Each column is obtained by taking an entity's motif vector and dividing by the total number of motifs for that entity, relative to the motif split in the timestamp-shuffled model. The entities are displayed left to right in descending order of total motif counts.}
    \label{fig:top10-fractions}
\end{figure*}

Among the top 10 nodes in each dataset (ordered by total motif counts),~\cref{fig:top10-fractions} shows that there are a number of distinct `signatures' which may signify different functional roles of the nodes in this network. One of the patterns (e.g. shared by nodes 3,6,8,9 and 10 in Alphabay and 8 in Hydra) is mostly having just pairwise incoming motifs, another (nodes 1, 4, 5 in Alphabay and 1, 2, 4 in Hydra) being star motifs with all outgoing edges.  The Alphabay node itself, despite its central role in the network as an escrow, placed 17th in total number of 1-hour motifs. The top node in Alphabay has 19 trillion motifs in total, a slightly negative transaction balance (about 2\% of its incoming transaction value) and a large out-degree (1831183) compared to in-degree (668). By checking with a blockchain explorer the addresses involved in the transaction and referencing with the GraphSense wallet/address labelling service~\cite{haslhofer2021graphsense}, these addresses seem to be associated with an illicit service that was very popular for a short amount of time.

Among the highest motif nodes in the NFT dataset, some (e.g. nodes 4, 5, 7, 9 and 10) show similarities to top nodes in Alphabay and Hydra with all-outgoing patterns being common, but we also see more incoming-based star motifs (nodes 6, 8) and the top node has more of a variety of motifs.

\subsection{Motifs reveal events not observable by studying transaction volume}
\label{sec:motifs-over-time}
\begin{figure*}[htbp]
  \begin{subfigure}[t]{.48\textwidth}
    \centering
    \includegraphics[width=\linewidth]{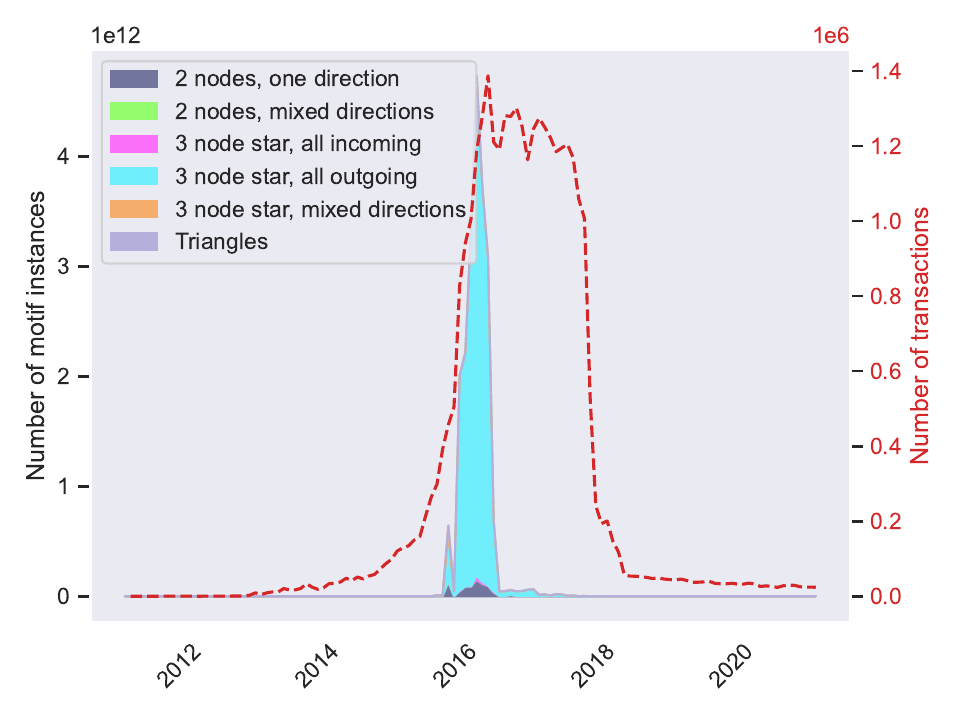}
    \caption{Monthly graph-wide motif counts over time for the Alphabay dataset separated by type. The dashed line shows the monthly number of transactions.}
  \end{subfigure}
  \hfill
  \begin{subfigure}[t]{.48\textwidth}
    \centering
    \includegraphics[width=\linewidth]{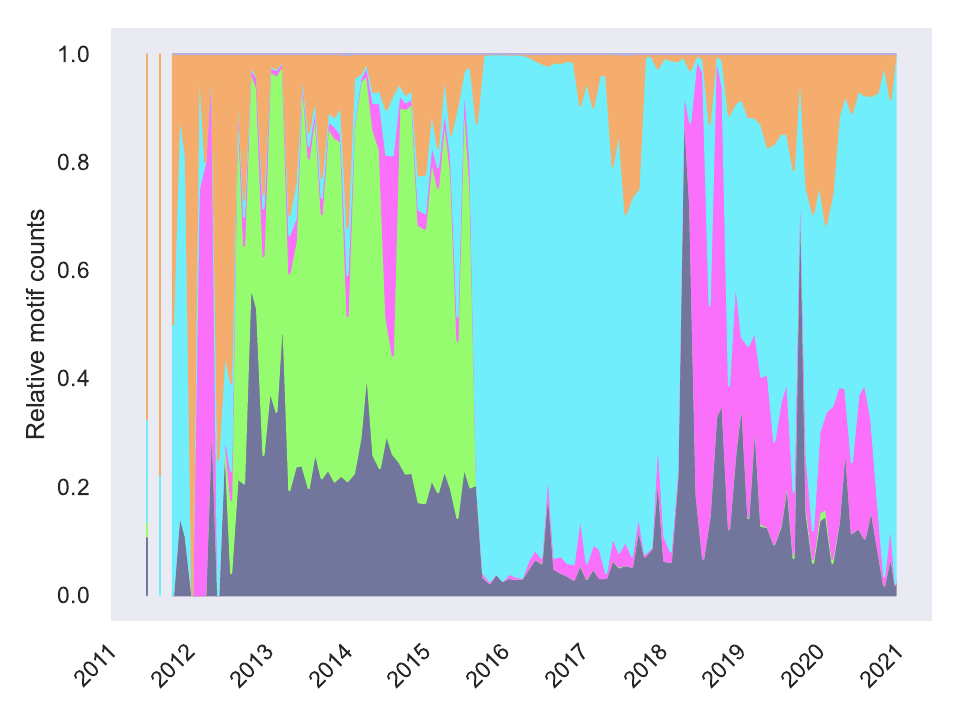}
    \caption{The monthly counts of different motif types in Alphabay as a proportion of all motifs. The classification of motifs and colour coding is the same as that used in (a). }
  \end{subfigure}
  \medskip
  \begin{subfigure}[t]{.48\textwidth}
    \centering
    \includegraphics[width=\linewidth]{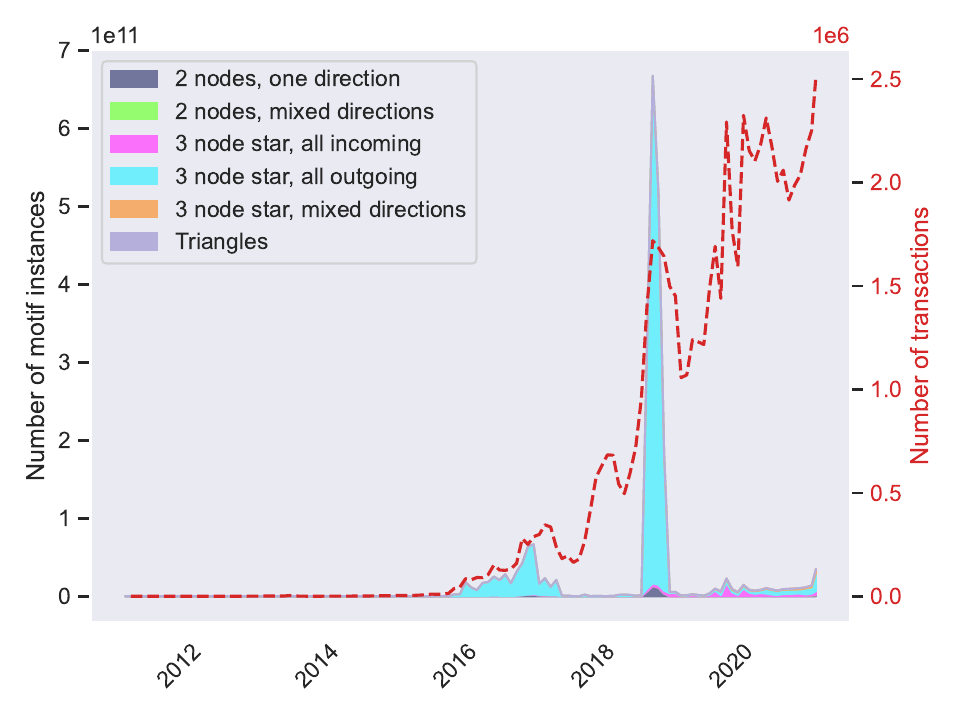}
    \caption{Monthly graph-wide motif counts over time for the Hydra dataset separated by type. The dashed line shows the monthly number of transactions.}
  \end{subfigure}
  \hfill
  \begin{subfigure}[t]{.48\textwidth}
    \centering
    \includegraphics[width=\linewidth]{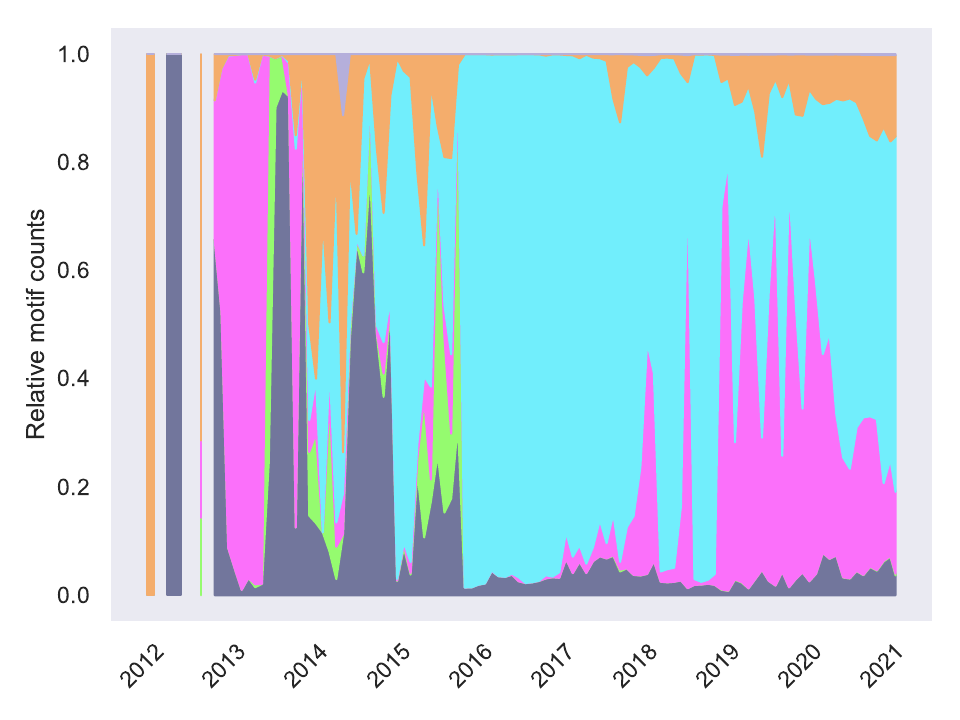}
    \caption{The monthly counts of different motif types in Hydra as a proportion of all motifs. The classification of motifs and colour coding is the same as that used in (a). }
  \end{subfigure}
  \hfill
\begin{subfigure}[t]{.48\textwidth}
    \centering
    \includegraphics[width=\linewidth]{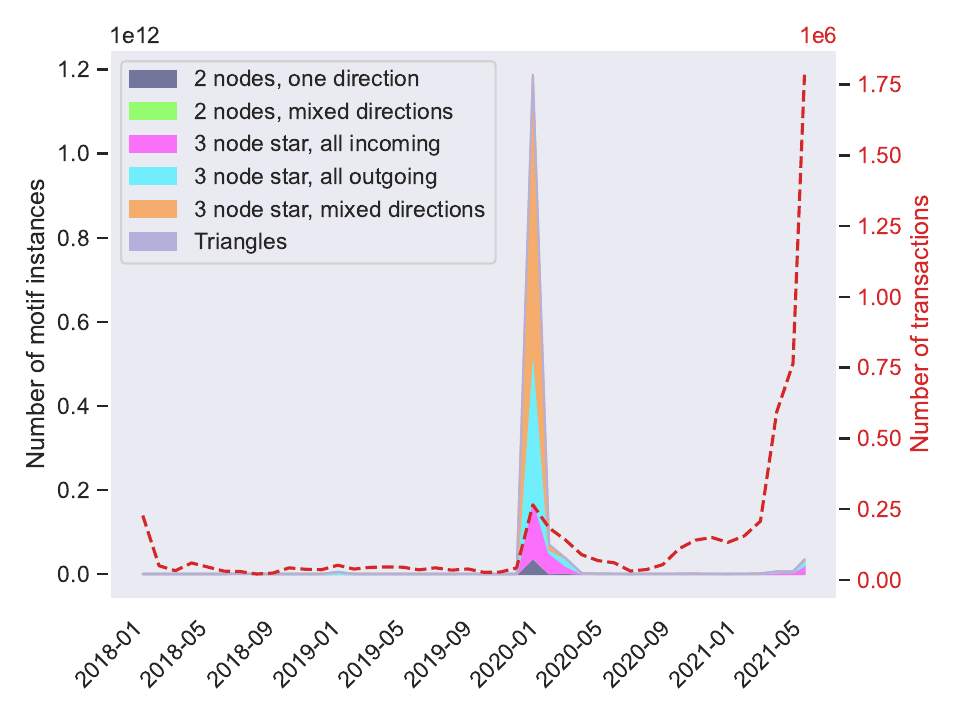}
    \caption{Monthly graph-wide motif counts over time for the NFT trade dataset separated by type. The dashed line shows the monthly number of transactions.}
  \end{subfigure}
  \hfill
  \begin{subfigure}[t]{.48\textwidth}
    \centering
    \includegraphics[width=\linewidth]{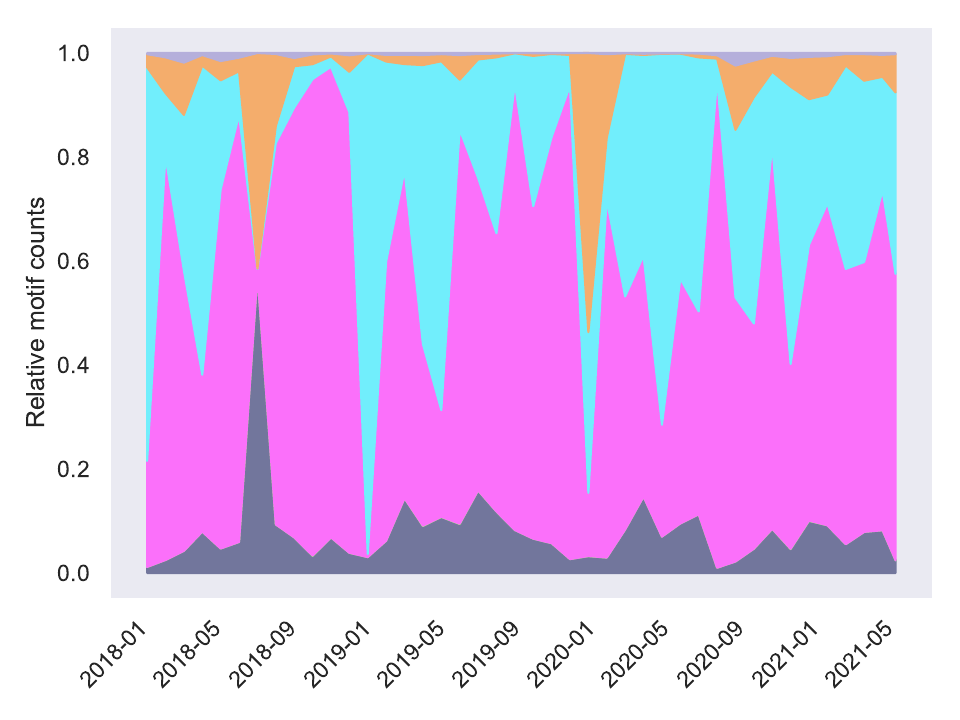}
    \caption{The monthly counts of different motif types in the NFT market as a proportion of all motifs. The classification of motifs and colour coding is the same as that used in (a). }
  \end{subfigure}
  \caption{Results on 1-hour motif counts for both datasets on a month-by-month basis. The left hand plots show the total monthly motif counts for each dataset grouped by motif types; the right hand shows these counts as a proportion of all motifs present.}
  \label{fig:monthly-motifs}
  \end{figure*}

Next we study how the motif counts are spread out through the time periods of the different datasets. To do this, we group the transactions by month and count the temporal motifs within that month with $\delta$ set to one hour for Alphabay and Hydra and $\delta$ set to one day for the NFT sales following previous experiments. \Cref{fig:monthly-motifs} shows these results, in terms of total motif counts and relative counts. In the Alphabay data, we see a spike in motif counts some months after the market opens in 2014, which at first seems to follow the trend in transaction volume. However, these counts drop off long before the transactions do, and long before the market is shut down. Given that the entity with the most motifs is possibly an illicit service or mixer (discussed in \cref{sec:localmotifs}), the drop-off could be due to users moving away from using it. In the Hydra market, we observe shadows of Alphabay's active period between 2014-2017 (some of Hydra's userbase were Alphabay users who migrated when it was shut down~\cite{nadini2022emergence}. This is followed by a similar spike and premature drop-off of motifs to Alphabay that is not mirrored in the transaction volume, which is still increasing at the end of our observation period. The vast majority of the NFT motifs occur within early 2020 when the technology was taking off followed by another spike at the end of the study period.

Following this, we investigate the composition of motifs over time, that is the counts of each category of motif as a proportion of all motif counts. In the Alphabay dataset, most of the motifs between 2013 and 2016 are two-node motifs representing transactions between peers (it is unlikely that these two-node motifs would involve the marketplace at this point as this would additionally invoke star motifs). At some point coinciding with the spike observed in all motif counts, this behaviour transitions to being all-outgoing stars which dominate, which could be the result of entities like the marketplace itself or an exchange facilitating transactions, or switching behaviour of users making purchases among different vendors. However, instead of the two-node motifs returning after Alphabay is shut down in 2017, the star motifs are still the most prevalent, fluctuating between all-incoming and all-outgoing. In the NFT dataset, apart from early 2020 where there is a spike in motif volume, most of the motifs are all-incoming stars. This could be due to the practice of sellers auctioning a number of their NFTs with the same closing period (and hence the transactions from winning bidders come in at the same time).

  \medskip

\subsection{Choosing and examining the effects of time period $\delta$}
\label{sec:timescale}
\begin{figure*}[htpb]
    \centering
    \begin{subfigure}[a]{0.9\textwidth}
        \centering \includegraphics[width=\linewidth]{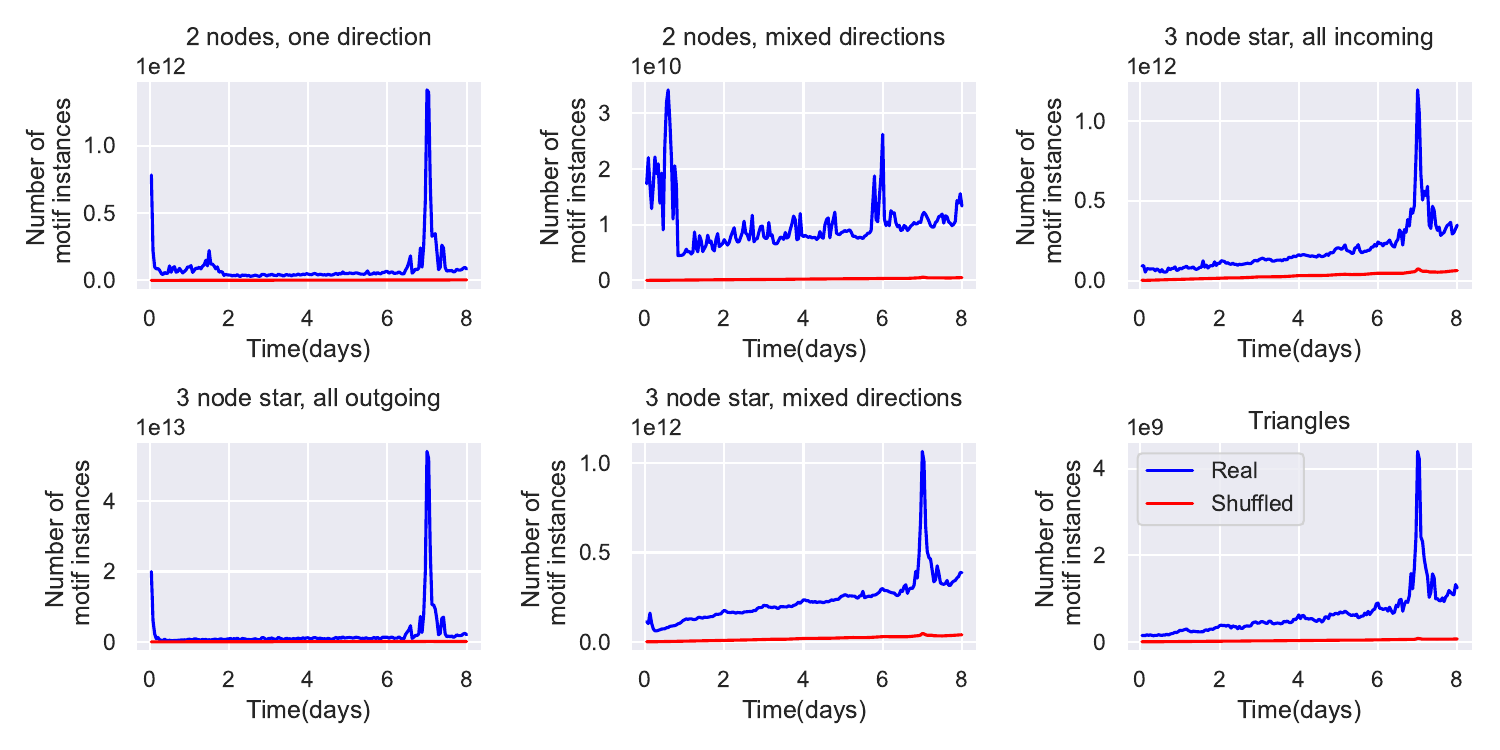}
        \label{fig:alpha-timescale}
        \caption{Alphabay transactions}
    \end{subfigure}
    \begin{subfigure}[b]{0.9\textwidth}
        \centering \includegraphics[width=\linewidth]{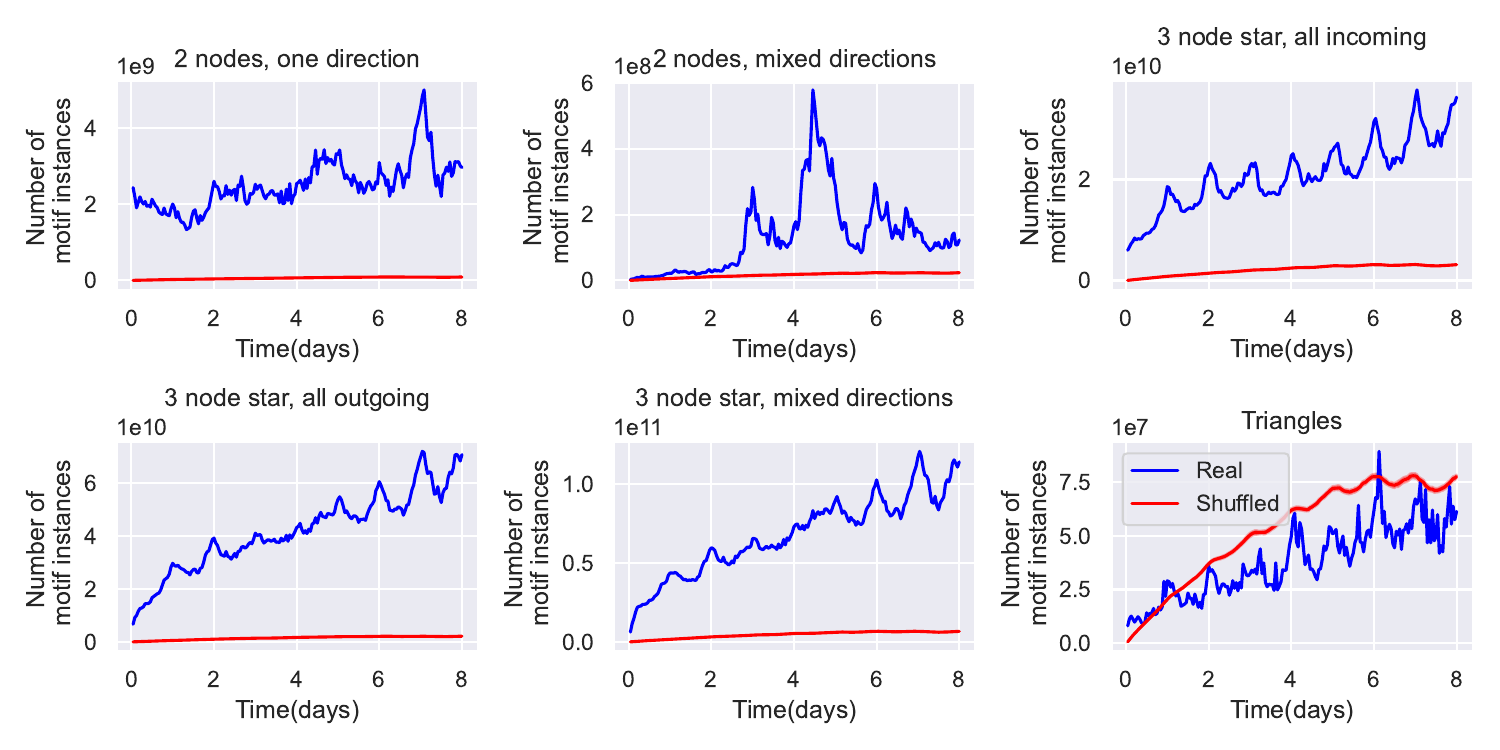}
        \label{fig:nft-timescale}
        \caption{NFT purchases}
    \end{subfigure}
    \caption{The number of each type of motif as $\delta$ the timescale is increased a single hour at a time for (a) Alphabay (b) NFT purchases. A peak in the graph at a timescale of $t$ hours shows that a large number of motifs are introduced by moving the timescale from $t-1$ to $t$ hours. In each graph, the blue line shows the count for the real dataset and the red line shows the mean value for the timestamp shuffled data calculated over 10 realisations, with standard deviation error bars.}
    \label{fig:timescale}
\end{figure*}

Obviously, the choice of $\delta$ the time period is important. In this section we justify the choice of one hour for Alphabay and Hydra and one day for NFT. We do this by asking how many extra motifs are introduced by increasing the time window by a single hour. This highlights time periods of particular importance. By subtracting the motifs obtained with time window $\delta$ plus one hour from those with time window $\delta$ we can investigate the new motifs introduced as the time window used is increased. Moreover we can assess the robustness of the $\delta$ parameter as a measure of the number of motifs. \Cref{fig:timescale} shows the effect of increasing $\delta$ from one hour up to just over one week (168 hours) by one hour increments. First, we see that for Alphabay, our chosen size for most of the analysis $\delta=1$ hour captures the vast majority of short-range motifs, seen by the dip from one to two hours across the most common motifs (those which are three nodes outgoing and three nodes mixed directions). Apart from the two nodes mixed directions motif, the increase in motifs is then stable up to seven days. The graph for two nodes mixed directions shows a variety of peaks at between one hour and twelve hours; the reason for this is not completely certain but these could be human-moderated trades between two individuals done on the same day. Finally, all motifs show a peak at one week, this is due to an autofinalisation escrow mechanism in Alphabay where trades are cleared if not confirmed by the buyer within seven days. This leads to a peak at precisely seven days where a transaction directly causes another exactly seven days later. This is not a particularly useful timescale for analysis though, firstly because it is very long and would lead to extremely large motif counts but mainly because the motifs are a result of a ``refund" mechanism rather than genuine trades. 

For the NFT trading the three node motif counts all show large peaks at intervals of one day. The peaks increase in size as the number of days increases. A possible explanation is that the motif counts are related to the human diurnal cycle and people tend to trade with people in the same geographic area. Whatever the mechanism, this makes $\delta$ of 1 day the most obvious time period to study. In the case of motifs involving only two nodes there is no clear peak in the motif counts but these motifs are relatively rare compared with three node motifs (less than 1\% of the number of three node motifs). It is also interesting to note that the time-shuffled null model produces as many or more triangles than the real data showing that triangular trade patterns between three individuals are rare in the NFT marketplace within the one week time scale. 

\section{Discussion and conclusion}
This paper investigated the use of temporal network motifs to explain trade in cryptocurrency networks. We showed how to optimise the time period $\delta$ (the maximum time for a temporal motif) and how varying it gave insight into the behaviour of the studied systems. Efficient temporal graph computation in the Raphtory software allowed us to investigate multiple values of $\delta$ and multiple time windows. This was impossible using existing bespoke temporal motifs software. It allowed us to identify behaviours we that we would not have been able to see from a single or small number of $\delta$ values. We could see both human (diurnal) and mechanism (e.g. escrow timeouts) induced behaviours in the graphs. 

In all three datasets a naive analysis of graph-wide temporal motif counts would be misleading. It is vital to consider the distribution of motifs over the time dimension and individual nodes (with particular attention to ``heavy-hitters"). All the networks studied had extreme nodes that contributed a high proportion of the motifs. Simply counting motifs for the Alphabay network, for example, would not tell you that a high proportion of all motifs were from a single node in a short (six month) period. For the networks with observations centered around dark markets (Alphabay and Hydra) it was unexpected that the nodes representing the markets themselves did not top the list of nodes by motif count. 

In the Hydra network we could clearly see evidence of synchronisation between multiple nodes likely indicating coordinated, automated behaviour. The top ten nodes by motif count could be grouped into distinct classes: mainly buying behaviour (multiple transfers to one or more nodes); mainly selling behaviour (multiple transfers from one or more other nodes) and a small number exhibiting a mixture of sell/buy behaviours. Triangular motifs were rare in all cases and lower than the null model (chance) rate for the NFT data. Done with care this form of analysis of networks leads to insights into trade behaviours not obtainable from other analysis tools. 

Given the diverse motif profiles of heavy-hitting users and the ability to uncover events not observable by simply counting transactions, we believe local temporal motifs may prove useful as features for classification or clustering tasks, such as in the domains of anti-money laundering or fraud detection. 

\section{Methods}
In this paper, we study three-edge, up-to-three node, temporal motifs, defined in~\cref{sec:motif-definition}. Three is chosen for the following reasons. Firstly, it lies in the middle of a tradeoff between descriptive power and computational complexity; there are only six possible two-edge temporal motifs but an algorithm for counting four-edge, up-to-four node, temporal motifs is not feasible to run on this scale of data because of the combinatorial explosion. In addition, with so many possible motifs interpretability becomes a problem. Secondly, a number of datasets have been studied using this technique~\cite{paranjape2017motifs,liu2022temporal,liu2023temporal}, hence our work is more comparable with that of other researchers.

\subsection{Counting local and global motifs}
In this paper, as well as studying global motif counts over the whole graph, we study motif counts from the perspective of the nodes involved in them. We refer to these as the \emph{local motif counts} and we also study them as they change in time. We first define a temporal graph $G = (V,T)$ where $V$ is a set of \emph{vertices} and $T$ is a sequence of tuples $(u_i, v_i, t_i)$, $1\leq i \leq m$ with $u_i, v_i \in V$ and $t_1 \leq t_2 \leq \dots, \leq t_{m-1} \leq t_m$ which we will refer to as \emph{temporal edges}, representing an event (e.g. a trade of currency) from node $u_i$ to node $v_i$ at time $t_i$.  In our data the event often also has a weight $w_i$ representing the amount of currency involved in the transaction. The local motif count of vertex $u$ for a given motif $M$ (motif definition in~\cref{sec:motif-definition}) refers to the number of times vertex $u$ has participated in motif $M$. To count the local motifs, we implement a modified version of the algorithm by Paranjape et al~\cite{paranjape2017motifs} in our temporal graph library Raphtory~\cite{steer2020raphtory,steer2024raphtory} which performs the counts in parallel across the vertices. Raphtory was built from the ground up to study large-scale temporal networks efficiently, including the possibility to use `windowed' approaches which enabled the study of temporal motifs over time (\cref{sec:motifs-over-time}). Global temporal motifs counting code from SNAP~\cite{leskovec2016snap} was tested on the Alphabay dataset but didn't complete due to out-of-memory issues, whereas the implementation in Raphtory took 3 minutes on the same hardware (2021 MacBook Pro with 16GB RAM). Note that our motif counting algorithm is fundamentally the same; in both implementations, motifs are counted for each vertex using the Paranjape et al counting algorithm~\cite{paranjape2017motifs}. In their work, these local counts are then summed to obtain graph-wide motif counts whereas we also study the local counts. Differences in runtime/memory usage are therefore likely due to differences in the underlying graph storage/low level functions of SNAP and Raphtory. With this in mind, we point the readers to their work~\cite{paranjape2017motifs} for pseudocode of the algorithm. We instead focus here on the details of what is meant by a local motif in terms of which nodes in a motif count that motif. A speed comparison with~\cite{paranjape2017motifs} is in~\cref{sec:speed}.
\subsubsection{Two-node motifs}
There are eight ($2^3$) possible two-node, three-edge, motifs when measured from the perspective of an individual node and this corresponds to three links in temporal order with two possible directions. When viewed globally, this collapses to the four possibilities shown in the bottom left of~\cref{fig:motif-types} when not viewed from the perspective of a single node. Consider the following example: from the point of view of a node pair $(u,v)$ it makes sense for the node $u$ to count a three edge temporal motif with $v$ that as ``outgoing, incoming, outgoing" and from the perspective of $v$ this same motif would be counted as ``incoming, outgoing, incoming". From a global perspective this is a single motif with transactions in alternating directions. For plotting, we often group these into all-outgoing, all-incoming and mixed directions. The time-complexity of counting two-node motifs is $O(E)$ where $E$ is the number of temporal edges.
\subsubsection{Three-node stars}
The star motifs (all motifs in~\cref{fig:motif-types} excluding the two-node motifs and triangle motifs) involve one central node having transactions with two other leaf nodes. In our work, we include an instance of this motif in a node $u$'s local motif count only if $u$ is the central node of this motif. One might argue that the two leaf nodes should also count this motif instance. However, we argue that (i) the central node is more of interest in such a motif, (ii) having all three nodes count the motif would result in more noise, and (iii) would be more computationally intensive to run, as it may instead involve running the global motifs count across the two-hop neighbourhood of each node. As with two-node motifs, the time-complexity of counting star motifs is $O(E)$.

It is useful to aggregate these star motifs into different sub-categories of all-outgoing, all-incoming and mixed-direction stars since they represent different behavioural patterns: it is more likely to see three outgoing transactions happen at a similar time than three incoming transactions since they originate from the same source, and mixed direction stars may involve some dependent sequence of events. This classification is used in some of the figures for clarity instead of showing all 36 global/40 local motifs.
\subsubsection{Triangles}
Triangle motifs are counted locally for each of the nodes involved in a triangle since, for each of those motifs, it is hard to make an argument for one of its constituent nodes being the most important. Additionally, there is little extra difficulty counting triangle motifs for each node (compared to leaf nodes of the star motifs for example), they are simply over-counted once for each node, with respect to the global counting algorithm which counts each triangle once. The time-complexity of counting triangle motifs in Raphtory is $O(VE + \tau E)$ where $\tau$ is the number of static triangles in the graph. The first term relates to the enumeration of static triangles in the graph, the second to counting the number of triadic temporal motifs given the edges involved in each static triangle. We argue that in cryptocurrency transaction networks, the first term tends to dominate since the number of triangles in such a network is fairly small; this may not be the case in social networks.

\subsection{Choice of timescale}
\label{sec:which-timescale}
Naturally, the choice of timescale $\delta$ plays a role in what motifs are discovered. Human communication patterns have been known to have characteristic timescales~\cite{aoki2016input} based on memory/attention~\cite{williams2022shape} and diurnal behaviour~\cite{arnold2021moving}. In transaction networks, we may also expect to see timescales resulting from pre-programmed rules such as preset durations for NFT auctions, transaction clearing times or payroll. While we perform a deeper study of which motifs arise at which timescale in~\cref{sec:timescale}, for most of the results we choose a timescale of one hour for the Alphabay/Hydra datasets and one day for the NFT datasets, corresponding to the first peak for a number of the motifs in~\cref{fig:timescale}. One observation which may be useful to practitioners is that one can study motifs which have a duration between values $\delta_1$ and $\delta_2 > 1$ by subtracting the motifs obtained with $\delta=\delta_1$ from those obtained with $\delta=\delta_2$. In some scenarios for example, motifs from transactions that all occur within the same block may be indicative to an algorithmic mechanism. If it is preferable to exclude such motifs in an effort to exclude pre-programmed behaviour, this can therefore be done by subtracting motifs with $\delta$ set to some time period much less than the block processing time e.g. 1 second.

To encourage researchers to explore this important effect of timescale choice on results using experiments like~\cref{fig:timescale}, our implementation of the motif mining algorithm in Raphtory includes the possibility to provide a range of delta values as input for the algorithm. This makes use of the observation that enumerating the static triangles in the network is typically the bottleneck of the algorithm's runtime. Having identified the edges which participate in these static motifs, running the motif counting procedure on these edges for each delta value becomes significantly faster than re-running the full algorithm for each delta value.

\subsection{Speed of execution compared to state-of-the-art}
\label{sec:speed}

\begin{figure}[htbp]
    \centering
    \includegraphics[width=0.6\linewidth]{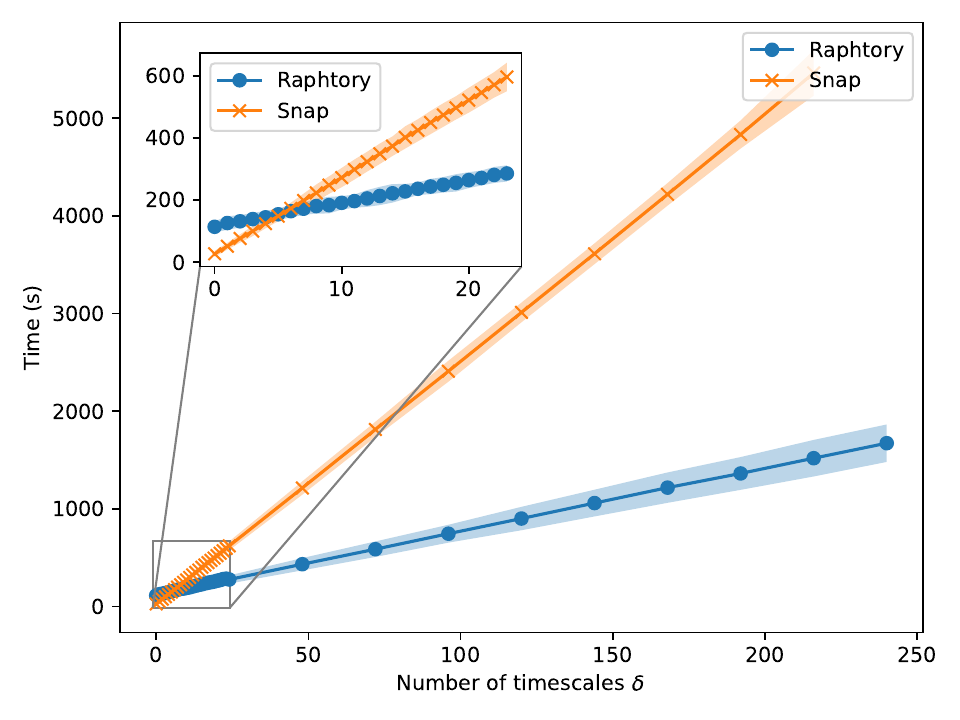}
    \caption{The time taken to count motifs for a range of $\delta$ values. For each value $\delta$ on the $x$-axis, the corresponding $y$ value is the time taken to count motifs for the input array $\left[1 \text{ hour}, 2 \text{ hours}, \dots , \delta \text{ hours}\right]$. The result is averaged over 10 experiments with shaded 95\% confidence intervals.}
    \label{fig:snap-raphtory}
\end{figure}

\Cref{fig:snap-raphtory} shows the time to count motifs for a range of $\delta$ values in the Raphtory implementation compared with what is arguably the most competitive alternative, the SNAP temporal motifs implementation. For each $\delta$ value on the $x$-axis, the global motifs algorithm is run for an array of timescales: 1 hour, 2 hours, 3 hours, ... , $\delta$ hours with the corresponding value on the $y$ axis showing the total runtime in seconds. Therefore, here $\delta$ represents both the number of timescales processed and the size of the largest timescale in hours.  For fewer than 6 windows, SNAP is faster, which is likely because Raphtory as a more general-purpose library incurs a slightly higher initial overhead for building the graph. However, the time increment for adding more windows in Raphtory is much smaller than the alternative in SNAP which can be seen by Raphtory's smaller slope. Note that in both implementations, the time is linear in the number of $\delta$ values processed and that the effect of the size of timescale $\delta$ itself on the time taken to process is negligible.

\subsection{Null model}
\label{sec:nullmodel}
To investigate whether motifs arise from temporally correlated behaviour or are simply a result of a large number of transactions we use a null model. In this case we use a randomised reference model where for our set of transactions $(u_i, v_i, t_i)$ the nodes $u_i$ and $v_i$ are kept the same and the $t_i$ are shuffled randomly among the transactions. This has the effect that between any node pair the same absolute number of transactions occurs over the whole study period and for any chosen time period the same number of transactions occurs in the network. We will refer to this model as the \emph{timestamp-shuffled} model~\cite{gauvin2022randomized}. For this paper, we produce 10 shuffled versions of each dataset and use the term ``relative to the null model" to mean that we divide the value as found in the real dataset by the mean of the values found on the 10 shuffled versions for that dataset.
\subsection{Transactions where time order is uncertain}
\label{sec:timeorder}
In a blockchain setting, transactions are released in \emph{blocks} according to when they are successfully verified by a miner, typically in time intervals of around 8-10 minutes. This means that while a transaction can be initiated by a user/service at any continuous time, it is only possible to observe the block-wide timestamp of each timestamp. The order in which they occur in the block is the order in which they were verified by a miner which comes with a topological order guarantee that if transaction $\tau_2$ spends an output of $\tau_1$ then $\tau_1$ must come before $\tau_2$ in the block~\cite{nakamoto2008bitcoin}, however there is no other guarantee on time ordering within the block. In fact, miners are incentivised to verify the transactions with the highest transaction fee first. For this work we take the block order to be the order of the transactions, but show in our supporting information that randomising intra-block orderings causes a deviation in the motif counts with a maximum at around 3\%.
\subsection{Data and code availability}
\label{sec:datacode}
Each of the algorithms (a local motif counting algorithm, a global motif counting algorithm, and a global motif algorithm optimised for multiple timescales) has been implemented in the open-source library Raphtory~\cite{steer2024raphtory} in Rust and Python. The Alphabay and Hydra datasets are proprietary but the NFT dataset is publicly available~\cite{nadini2021mapping} and scripts for reproducing the NFT results can be found at \url{https://github.com/narnolddd/motif-paper-reproduce}. All plots were generated using Matplotlib 3.8.2 (\url{https://matplotlib.org/stable/}) and Seaborn 0.13.2 (\url{https://seaborn.pydata.org/}).
\section*{Additional information}
The authors declare no competing interests. Thank you to Chainalysis for providing two of the datasets used in this research.
\bibliographystyle{abbrv}
\bibliography{motifs_bib}

\end{document}